\documentclass{article}
\usepackage[utf8]{inputenc}
\usepackage{authblk}
\usepackage{setspace}
\usepackage[margin=1.25in]{geometry}
\usepackage{graphicx}
\graphicspath{ {./figures/} }
\usepackage{subcaption}
\usepackage{amsmath}
\usepackage{lineno}

\usepackage[style=nejm, 
citestyle=numeric-comp,
sorting=none]{biblatex}
\title{Magnetic Fusion Plasma Drive}

\author[1,2*]{Florian Neukart}

\affil[1]{Leiden Institute of Advanced Computer Science, Leiden University, Leiden, Netherlands}
\affil[2]{Terra Quantum AG, St. Gallen, Switzerland}

\date{}

\begin{document}

\maketitle

\begin{abstract}
In the evolving realm of space exploration, efficient propulsion methods are paramount to achieve interplanetary and possibly interstellar voyages. Traditional propulsion systems, although proven, offer limited capabilities when considering longer-duration missions beyond our immediate cosmic vicinity. This paper introduces and thoroughly investigates the Magnetic Fusion Plasma Drive (MFPD) propulsion system, a novel fusion-powered propulsion mechanism. Through rigorous theoretical underpinnings and mathematical formulations, we elucidate the principles governing fusion reactions in the context of propulsion, plasma dynamics, and magnetic confinement in space. Comparative analyses indicate significant advantages of the MFPD system over existing technologies, particularly in fuel efficiency, thrust capabilities, and potential scalability. Example calculations further substantiate the immense energy potential and feasibility of the MFPD for long-duration missions. While challenges remain, the MFPD system embodies a promising avenue for a propulsion paradigm shift, potentially revolutionizing our approach to space exploration.
\end{abstract}


\section{Introduction}\label{Introduction}

The challenge of deep-space exploration and transporting significant payloads across interplanetary and interstellar distances necessitates developing efficient and powerful propulsion systems. As humanity contemplates the establishment of colonies on distant planets and mining celestial bodies, our spacecraft's propulsion mechanisms become increasingly critical.

\subsection{Background on Space Propulsion Needs for Large Spacecraft}

Large spacecraft designed for transporting significant payloads or for long-duration missions often face challenges in terms of propulsion. The propulsion system must maintain efficiency over extended durations, be scalable, and offer a balance between thrust and fuel consumption. While instrumental in our space exploration endeavors, the current propulsion methods have clear limitations when scaling up for larger spacecraft or missions aiming at deep space destinations \cite{Mattingly2016, Cassady2020, Griebel2018}.

\subsection{Shortcomings of Current Propulsion Methods}

Chemical propulsion, the mainstay for current spacecraft, offers high thrust but has a limited specific impulse, making it unsuitable for prolonged deep-space missions \cite{Mattingly2016}. Similarly, while ion and electric propulsion systems provide high efficiency, their low thrust capabilities pose challenges for the rapid transport of large spacecraft \cite{Gallimore2004}. Solar sails and Nuclear Thermal Propulsion, while promising, have their challenges, which will be detailed in subsequent sections \cite{Cassady2020, Palaszewski2008, Borowski2001}.

\subsection{Objective and Overview of the Proposed MFPD System}

This paper introduces the Magnetic Fusion Plasma Drive (MFPD), a propulsion concept that seeks to harness the immense energy potential of nuclear fusion combined with magnetically confined plasma to produce thrust. The MFPD aims to address the limitations of current propulsion systems by providing a balance between thrust and efficiency, all while ensuring scalability for larger spacecraft. To achieve this, we draw on research in nuclear fusion \cite{Chang2015}, plasma physics \cite{Stacey2010}, and magnetohydrodynamics \cite{Freidberg1987}. The ensuing sections will consider the intricacies of current propulsion systems, the theoretical basis for the MFPD, mathematical formulations describing its operation, and a comparative analysis of its potential advantages and challenges \cite{Mattingly2016, Cassady2020, Chang2015, Stacey2010}.

\section{Background on Existing Propulsion Systems}
The vastness of space demands propulsion systems that are both efficient and capable of delivering sustained thrust over extended durations. Historically, the domain of space propulsion has been characterized by a spectrum of technologies, each developed to address specific mission requirements and the inherent challenges of space travel. As we think about more ambitious missions, it becomes crucial to understand the underpinnings, advantages, and limitations of the current state-of-the-art propulsion technologies. This chapter provides a comprehensive overview of the primary propulsion systems that have been, and continue to be, pivotal in our space exploration endeavors. We begin with the conventional chemical propulsion systems, which, for decades, have been fundamental to our space ventures. The discourse then progresses to ion and electric propulsion systems, highlighting their role in providing fuel-efficient solutions for prolonged space missions. We touch upon the principles and prospects of nuclear thermal propulsion and its potential in bridging the gap between efficiency and high thrust. The narrative then delves into the innovative concepts of solar sails, fusion propulsion, and the theoretical domain of antimatter propulsion.

In understanding each system's principles, merits, and challenges, we aim to provide foundational knowledge that sets the stage for introducing our proposed Magnetic Fusion Plasma Drive in subsequent chapters.

\subsection{Chemical Propulsion}

Chemical propulsion remains the primary means of propulsion for most contemporary spacecraft and has enabled a wide range of missions, from satellite launches to interplanetary explorations. The principle behind chemical propulsion is the combustion of chemical propellants to produce high-temperature and high-pressure gases that are expelled through a nozzle, resulting in thrust via Newton's third law \cite{Sutton2001}.

\subsubsection{Basic Principles and Mechanics}

The basic operation of a chemical rocket can be summarized by the following steps:

\begin{itemize}
    \item Combustion of propellants in a combustion chamber produces high-energy gases.
    \item The rapid expansion of these gases is channeled through a nozzle.
    \item As the gases exit the nozzle at high velocities, a force is exerted on the rocket in the opposite direction, propelling it forward.
\end{itemize}

Eq. \eqref{eq:thrust} gives the thrust \( F \) produced by a rocket:

\begin{equation}
    F = \dot{m} V_e + (P_e - P_0) A_e
    \label{eq:thrust}
\end{equation}

Where:

\begin{itemize}
    \item \( \dot{m} \) is the propellant mass flow rate.
    \item \( V_e \) is the exhaust velocity.
    \item \( P_e \) is the exhaust pressure.
    \item \( P_0 \) is the ambient pressure.
    \item \( A_e \) is the nozzle exit area \cite{Sutton2001}.
\end{itemize}

The specific impulse \( I_{sp} \), a measure of rocket propellant efficiency, is defined as the thrust per unit weight flow rate of the propellant (Eq. \eqref{eq:specific_impulse}):

\begin{equation}
    I_{sp} = \frac{F}{g_0 \dot{m}}
    \label{eq:specific_impulse}
\end{equation}

Where \( g_0 \) is the standard gravitational acceleration \cite{Mattingly2016}.

\subsubsection{Limitations in terms of Specific Impulse and Fuel Mass}

While chemical propulsion offers significant thrust, allowing for rapid changes in velocity, its specific impulse values are inherently limited by the chemical propellants' energy content. Typically, chemical rockets have \( I_{sp} \) values in the range of 250 to 450 seconds for bipropellant systems \cite{Mattingly2016}. This limitation implies a substantial fuel mass is required for long-duration missions or missions requiring significant velocity changes.

Moreover, the exponential nature of the rocket equation (Eq. \eqref{eq:rocket_equation}):

\begin{equation}
    \Delta V = V_e \ln\left(\frac{m_0}{m_f}\right)
    \label{eq:rocket_equation}
\end{equation}

Where:

\begin{itemize}
    \item \( \Delta V \) is the change in velocity.
    \item \( V_e \) is the effective exhaust velocity.
    \item \( m_0 \) is the initial total mass.
    \item \( m_f \) is the final total mass \cite{Mattingly2016}.
\end{itemize}

emphasizes the challenges of achieving high \( \Delta V \) maneuvers. As the required \( \Delta V \) for a mission increases, the ratio of initial to final mass becomes increasingly larger, necessitating an even greater propellant mass. 

This inherent limitation restricts the feasible mission profiles for spacecraft reliant solely on chemical propulsion, especially for deep-space endeavors. The significant mass and volume of propellant needed can make some missions infeasible or require complex mission architectures involving multiple launches and in-space assembly or refueling \cite{Mattingly2016, Cassady2020}.

\subsection{Ion and Electric Propulsion}

Ion and electric propulsion systems have gained popularity for their high efficiency, especially in long-duration missions where their prolonged thrust capability can achieve substantial velocity changes over time \cite{Choueiri2009}. Unlike chemical rockets, which rely on the combustion of propellants, ion and electric thrusters use electricity (often from solar panels) to ionize a propellant and accelerate it using electromagnetic fields.

\subsubsection{Mechanisms of Ion/Electric Propulsion}

Electric propulsion can be categorized based on the method of accelerating the propellant:

\begin{itemize}
    \item Electrothermal Thrusters: These thrusters heat the propellant using electrical power, which then expands and is expelled through a nozzle to produce thrust \cite{Mattingly2016}.
    \item Electrostatic Thrusters (e.g., Hall Effect Thrusters and Gridded Ion Thrusters): Propellant atoms are ionized and then accelerated using electric or magnetic fields. Electrons then neutralize the positively charged ions upon exit to produce a neutral exhaust \cite{Choueiri2009}.
    \item Electromagnetic Thrusters (e.g., Magnetoplasmadynamic Thrusters): These use both electric and magnetic fields to accelerate the ionized propellant \cite{Mattingly2016}.
\end{itemize}

The exhaust velocity of the ionized propellant predominantly determines the specific impulse of electric propulsion systems. The relationship is given by Eq. \eqref{eq:isp_electric}:

\begin{equation}
    I_{sp} = \frac{V_e}{g_0}
    \label{eq:isp_electric}
\end{equation}

Where \( V_e \) is the exhaust velocity and \( g_0 \) is the standard gravitational acceleration \cite{Mattingly2016}.

The specific impulse, \( I_{sp} \), is a key parameter used to characterize the performance of a propulsion system. It's essentially the impulse (change in momentum) provided per unit of propellant mass expended. The impulse provided by a thruster is given by \( F \times dt \), where \( F \) is the thrust and \( dt \) is an infinitesimal duration. The propellant mass expended in this duration is \( \dot{m} dt \). Thus, the specific impulse, by definition, is given by Eq. \eqref{eq:specific_impulse_ref}:

\begin{equation}
    I_{sp} = \frac{F dt}{g_0 \dot{m} dt} = \frac{F}{g_0 \dot{m}}
    \label{eq:specific_impulse_ref}
\end{equation}

which is as per Eq.~\eqref{eq:specific_impulse}.

However, the thrust, \( F \), generated by a propulsion system is also related to the exhaust velocity of the expelled propellant, \( V_e \), by the equation Eq. \eqref{eq:thrust_relation}:

\begin{equation}
    F = \dot{m} V_e
    \label{eq:thrust_relation}
\end{equation}

Substituting the value of \( F \) from Eq. \eqref{eq:thrust_relation} into Eq.~\eqref{eq:specific_impulse_ref}, we get Eq. \eqref{eq:isp_electric}. 

Hence, while Eq.~\eqref{eq:specific_impulse} relates \( I_{sp} \) to the thrust and propellant mass flow rate, Eq.~\eqref{eq:isp_electric} provides a direct relation to the exhaust velocity, which is often more convenient when analyzing electric propulsion systems, where the exhaust velocities can be exceptionally high \cite{Choueiri2009,sutton2001rocket}.

\subsubsection{Limitations in Thrust Capabilities}

While electric propulsion systems excel in terms of specific impulse (often achieving \( I_{sp} \) values in the range of 1000 to 5000 seconds or higher), they typically have low thrust levels compared to chemical systems. This means they impart changes in velocity over longer durations, making them unsuitable for applications requiring rapid thrust maneuvers, such as ascent or landing on celestial bodies.

Moreover, power generation and heat dissipation become critical issues. The efficiency of electric propulsion is closely tied to the available electrical power. As the power requirements increase, so does the need for large solar panels or nuclear power sources, potentially increasing the spacecraft's mass and complexity \cite{Mattingly2016}.

Another challenge is the erosion of thruster components due to the high-energy ionized propellant. Over extended missions, this can reduce thruster lifespan and performance. Additionally, the need for precise propellant ionization and acceleration mechanisms makes electric propulsion systems more complex and potentially susceptible to technical malfunctions \cite{Choueiri2009}.

Furthermore, spacecraft employing electric propulsion systems often follow spiral trajectories, especially when transitioning in and out of gravitational wells. During these trajectories, a portion of the thrust is continuously expended to counteract the gravitational pull of the celestial body, leading to gravity losses. Such losses can have a notable impact on mission duration and the efficiency of propellant usage. While high \( I_{sp} \) values do imply reduced propellant consumption, gravity losses can offset this advantage, especially during extended periods of thrusting in a gravitational field. This aspect can particularly affect missions that transition between different orbits around a planet or moon, where the spiraling trajectory is pronounced \cite{Choueiri2009,chen1984introduction}.

\subsection{Nuclear Thermal Propulsion (NTP)}

Nuclear Thermal Propulsion represents a distinct branch of propulsion technology that capitalizes on the energy released from nuclear reactions, specifically nuclear fission, to heat a propellant and generate thrust. Historical interest in NTP emerged during the Cold War era, with notable projects such as the U.S.'s Project Rover. Though not currently in widespread use, NTP holds potential for future deep space missions due to its promise of high thrust combined with relatively high specific impulse \cite{Borowski2001}.

\subsubsection{Mechanism and Operation of NTP}

The fundamental operation of an NTP system can be described as:

\begin{itemize}
    \item A nuclear reactor, containing fissile material, initiates a controlled nuclear fission reaction, releasing a substantial amount of thermal energy.
    \item A propellant (commonly hydrogen) is passed through the reactor core, where it gets heated to high temperatures by the thermal energy from the fission reactions.
    \item The heated propellant expands and is expelled through a nozzle, producing thrust in a manner analogous to a chemical rocket.
\end{itemize}

Unlike chemical propulsion, where the energy source and the propellant are the same, NTP decouples the energy source (nuclear reactor) from the propellant (hydrogen). This allows for a higher specific impulse because the exhaust velocities can be much greater than those achievable with chemical reactions alone \cite{Borowski2001}.

\subsubsection{Advantages and Potential}

NTP systems can achieve specific impulse values between 850 to 1000 seconds, nearly double that of the most efficient chemical rockets. This offers a balance between the high thrust of chemical rockets and the high efficiency (in terms of specific impulse) of electric propulsion. Such a combination is particularly valuable for crewed missions to distant planets, where minimizing travel time is crucial \cite{Borowski2001}.

\subsubsection{Limitations and Challenges}

\begin{itemize}
    \item Radiation Concerns: One of the primary challenges with NTP is managing the radiation produced by the nuclear reactor. This necessitates robust shielding, both to protect spacecraft systems and, in the case of crewed missions, to ensure the safety of astronauts \cite{Borowski2001}.

    \item Technical Complexity: The need to handle fissile material, control nuclear reactions, and manage reactor temperatures presents considerable engineering challenges \cite{Borowski2001}.

    \item Environmental and Safety Concerns: The potential consequences of an accident, either during launch or in operation, have made NTP a contentious choice. The release of radioactive materials could pose environmental risks \cite{Borowski2001}.

    \item Political and Regulatory Hurdles: Deploying nuclear technology in space involves navigating a complex landscape of international treaties and regulations \cite{Borowski2001}.
\end{itemize}

\subsection{Solar Sails}

Solar sails, or photon sails, offer a radically different approach to propulsion in space. Instead of expelling mass to achieve thrust, solar sails harness the momentum of photons (light particles) emitted by the sun. As these photons reflect off the sail, they transfer momentum, generating a propulsive force. Though the force exerted by individual photons is minuscule, the cumulative effect over vast areas and extended durations can result in significant acceleration \cite{MacNeal2020}.

\subsubsection{Principle and Operation}

Solar sails operate on the principle of radiation pressure. When photons reflect off a surface, they transfer twice their momentum to that surface. For a perfectly reflecting sail oriented perpendicular to the sun, the force \( F \) due to radiation pressure can be given by Eq. \eqref{eq:solar_sail_force}:

\begin{equation}
    F = \frac{2 I A}{c}
    \label{eq:solar_sail_force}
\end{equation}

Where:

\begin{itemize}
    \item \( I \) is the solar radiation intensity, typically around \( 1361 \, {W/m}^2 \) near Earth.
    \item \( A \) is the area of the sail.
    \item \( c \) is the speed of light \cite{MacNeal2020}.
\end{itemize}

\subsubsection{Advantages and Potential}

\begin{itemize}
    \item Fuel-less Propulsion: Since solar sails do not rely on onboard fuel or propellant, they can continue to accelerate as long as they remain exposed to solar radiation. This offers the potential for long-duration missions without the need for fuel resupply \cite{Matloff2010}.
   
    \item Scalability: Larger sails capture more photons, resulting in greater thrust. Advances in materials science can lead to lightweight, yet large sails that can harness substantial radiation pressure \cite{Matloff2010}.
   
    \item Interstellar Potential: While slow to start, over incredibly long distances and timeframes, solar sails could achieve a significant fraction of the speed of light, making them a contender for interstellar missions, especially when paired with powerful lasers that act as beamed propulsion sources \cite{Matloff2010}.
\end{itemize}

\subsubsection{Limitations and Challenges}

\begin{itemize}
    \item Initial Slow Acceleration: The thrust provided by solar radiation is gentle, making solar sails unsuitable for rapid maneuvers or missions requiring swift velocity changes.
   
    \item Distance from Sun: As a spacecraft ventures farther from the sun, the solar radiation intensity diminishes, leading to decreased thrust \cite{Matloff2010}.
   
    \item Material and Manufacturing: Crafting large, ultra-thin, and durable sails that can endure the space environment is a significant engineering challenge \cite{MacNeal2020}.
   
    \item Control and Navigation: Steering a spacecraft using a solar sail requires precise control of the sail's orientation relative to the sun. Achieving desired trajectories involves continuously adjusting this angle \cite{Matloff2010}.
\end{itemize}

\subsection{Fusion Propulsion}

Fusion propulsion is a concept that envisions harnessing the immense energy released from nuclear fusion reactions to propel spacecraft. Unlike nuclear fission, which involves splitting atomic nuclei, fusion combines light atomic nuclei, typically isotopes of hydrogen, to form heavier nuclei. In the process, a tremendous amount of energy is released, surpassing that of any chemical reaction \cite{chen1984introduction}.

\subsubsection{Principle and Operation}

The fundamental operation of fusion propulsion can be described as:

\begin{itemize}
    \item Fusion reactions are initiated in a contained environment, often through the use of magnetic or inertial confinement methods.
    \item The high-energy particles and radiation produced in the fusion reactions are directed out of the spacecraft through a magnetic nozzle or other mechanism, generating thrust.
    \item Additional propellants, such as hydrogen, can be introduced and heated by the fusion reactions, producing additional thrust like nuclear thermal propulsion \cite{chen1984introduction}.
\end{itemize}

The energy release per fusion reaction is given by Eq. \eqref{eq:fusion_energy}:

\begin{equation}
    E = \Delta m \times c^2
    \label{eq:fusion_energy}
\end{equation}

Where:
\begin{itemize}
    \item \( \Delta m \) is the change in mass between the initial reactants and the final products.
    \item \( c \) is the speed of light \cite{chen1984introduction}.
\end{itemize}

\subsubsection{Advantages and Potential}

\begin{itemize}
    \item High Specific Impulse: Fusion propulsion can theoretically achieve specific impulse values exceeding those of both chemical rockets and nuclear thermal propulsion, making long-duration missions more feasible.
   
    \item Abundant Fuel Sources: The primary fuel for fusion, isotopes of hydrogen-like deuterium and tritium, can be found in water, making them relatively abundant in the universe \cite{chen1984introduction}.
   
    \item Reduced Radiation Concerns: Unlike fission, fusion does not produce long-lived radioactive waste, mitigating some radiation concerns associated with nuclear propulsion.

    \item Vast Energy Potential: A small mass of fusion fuel can produce tremendous energy, potentially allowing for rapid transits between distant celestial bodies. Notably, this refers to energy per unit mass; volumetrically, fission fuels release more energy per cubic meter of fuel \cite{chen1984introduction}.

\end{itemize}

\subsubsection{Limitations and Challenges}

\begin{itemize}
    \item Technical Complexity: Achieving the conditions for controlled fusion reactions is an immense engineering challenge. Despite decades of research on Earth, we have yet to achieve sustained and net-energy-positive fusion reactions.
   
    \item Heat Management: The temperatures associated with fusion reactions are extremely high, demanding advanced materials and systems to handle the generated heat.

    \item Fuel Availability: While deuterium is relatively abundant, tritium is rare and has to be bred from lithium or other processes. Other fusion fuels, like helium-3, may be rare on Earth but could be mined from celestial bodies like the Moon.

    \item Magnetic Confinement: Using magnetic fields to confine the hot plasma in fusion reactors poses power requirements and stability challenges.

    \item Safety and Containment: Ensuring the safe containment of fusion reactions, especially when control might be lost, is critical \cite{chen1984introduction}.

\end{itemize}

While fusion propulsion remains in the realm of future possibilities, its promise of efficient, high-energy propulsion drives continued interest and research. If the challenges associated with controlled fusion are overcome, it could revolutionize space travel, reducing transit times and expanding our reach within our solar system and beyond.

\section{Theoretical Basis for the Magnetic Fusion Plasma Drive (MFPD)}

Nuclear fusion is when atomic nuclei come together to form a heavier nucleus. This process releases vast amounts of energy, primarily because the mass of the resulting nucleus is slightly less than the sum of its constituents. The difference in mass is released as energy according to Einstein's equation \(E=mc^2\).

\subsection{Nuclear Fusion in Space Propulsion}

In space propulsion, the most commonly considered fusion reactions involve isotopes of hydrogen: deuterium (D) and tritium (T). The primary fusion reaction can be represented as Eq. \eqref{eq:fusion_reaction}:

\begin{equation}
    D + T \rightarrow ^4 He + n + 17.6 MeV
    \label{eq:fusion_reaction}
\end{equation}

Where:
\begin{itemize}
    \item \(^4 He\) is helium-4.
    \item \(n\) is a neutron.
\end{itemize}

This reaction releases 17.6 MeV (mega-electronvolts) of energy, predominantly carried away by the neutron.

\subsubsection{Advantages over fission and chemical reactions}

Fusion has several advantages as a propulsion mechanism:

\begin{itemize}
    \item Higher Energy Density (J/Kg): Fusion reactions release more energy per unit mass than chemical reactions or nuclear fission. Specifically, while a uranium fission reaction releases approximately 197 MeV of energy per fission, fusion reactions release significantly more energy per kilogram. It's imperative to note that this advantage is specifically regarding energy per unit mass (J/Kg). Regarding volumetric energy density \(J/m^3\), fission reactions can release more energy than fusion \cite{FusionEnergy2018}.
    
    \item Abundant Fuel: Deuterium can be extracted from seawater, and while tritium is rare, it can be bred from lithium, which is relatively abundant \cite{Dean2003}.
    
    \item Safety: Fusion doesn't suffer from the same meltdown risks as fission. Furthermore, while tritium is radioactive, it's less of a long-term contaminant than many fission by-products \cite{Hester2007}.
\end{itemize}

\subsection{Plasma Dynamics}

Plasma is often termed the fourth state of matter. It consists of charged particles: ions and electrons. In fusion propulsion, the fuel (like D-T mix) is heated to such high temperatures that it becomes plasma \cite{Chen1984}.

\subsubsection{Properties and behaviors of high-energy plasma}

\begin{itemize}
    \item Conductivity: Plasma is a good conductor of electricity due to the free ions and electrons \cite{Stacey2005}.
    
    \item Reactivity: High-energy plasma undergoes fusion reactions at sufficiently high temperatures and pressures \cite{FusionPrinciplesAndPlasmaPhysics2007}.
    
    \item Responsiveness to Magnetic Fields: Being charged, plasma responds strongly to magnetic fields, allowing for magnetic confinement \cite{Hutchinson2008}.
\end{itemize}

\subsubsection{Importance of magnetic confinement}

To achieve fusion, plasma must be confined at high temperatures and pressures for a sufficient duration. Magnetic confinement uses magnetic fields to contain the plasma, preventing it from coming into contact with (and being cooled by) the walls of the containment vessel \cite{FusionPrinciplesAndPlasmaPhysics2007}.

\subsection{Magnetic Confinement in Propulsion}
The magnetic confinement technique in propulsion systems serves as a revolutionary approach to harness plasma's immense power and potential, the fourth state of matter. Utilizing magnetic fields, it's possible to effectively control, guide, and confine the hot plasma, which is inherently challenging due to its high-energy nature and erratic behavior. These confinement strategies promise efficient plasma management and pave the way for innovative propulsion methods that could redefine space exploration.

\subsubsection{How magnetic fields can be used to control and direct plasma}

The charged particles in plasma follow helical paths around magnetic field lines. By carefully designing the magnetic field topology, one can ensure that plasma remains confined in a desired region \cite{Hutchinson2008}.

\begin{itemize}
    \item Tokamak Configuration: This doughnut-shaped configuration combines external magnetic coils and a toroidal current within the plasma to create a strong confining magnetic field \cite{Wesson2004}.
    
    \item Magnetic Nozzles: In the context of propulsion, magnetic fields can be shaped to form a "nozzle" that directs the high-energy plasma out of the thruster, generating thrust \cite{Ferrario2017}.
\end{itemize}

\subsubsection{The role of superconducting magnets}

Superconducting magnets are critical in advanced plasma confinement schemes because they produce strong magnetic fields with minimal energy consumption. These magnets can carry large currents without electrical resistance when cooled below their critical temperature. They're essential for large-scale, sustained magnetic confinement of plasma \cite{Schober2008}.

\subsubsection{Magnetic confinement in the MFPD}
The nomenclature "Magnetic Fusion Plasma Drive" indeed suggests that magnetic confinement plays a pivotal role in the system. Magnetic confinement's primary purpose in the MFPD is to contain and stabilize the plasma, ensuring that the fusion reactions occur efficiently. At its core, MFPD uses magnetic fields to initiate and sustain the fusion process. The initial state of the plasma can be achieved through various methods such as electromagnetic induction, radio frequency heating, or neutral beam injection, all of which directly or indirectly leverage magnetic fields. Once the plasma is heated to the necessary conditions for fusion, maintaining those conditions and preventing the plasma from interacting with the walls (and hence cooling down) becomes crucial. This is where magnetic confinement plays its most significant role. The term "generalized magnetic confinement approach" in the manuscript is intended to convey that while specific configurations like Tokamak, Stellarator, or Field-Reversed Configuration are well-known and widely studied, MFPD might employ a combination or a variation of these configurations to achieve the desired plasma conditions and confinement. The unique design considerations for a propulsion system (as opposed to a terrestrial power generation setup) might necessitate such an approach.

It's also important to note that while magnetic confinement is pivotal, additional stabilization mechanisms, both passive and active, might be employed to ensure the plasma's stability and longevity. Magnetic confinement for plasma control is one way to confine the plasma in the MFPD system. A foundational principle behind this confinement is the Lorentz Force experienced by charged particles moving in a magnetic field (Eq. \eqref{eq:lorentz_force}):

\begin{equation}
\vec{F} = q(\vec{v} \times \vec{B})
\label{eq:lorentz_force}
\end{equation}

Here, \( q \) is the charge of the particle, \( \vec{v} \) represents its velocity, and \( \vec{B} \) denotes the magnetic field \cite{griffiths2013introduction,jackson1998classical}.

Charged particles will gyrate around the field lines upon interaction with a magnetic field. The gyroradius or the radius of this spiral is given by Eq. \eqref{eq:gyroradius}:

\begin{equation}
r_g = \frac{mv_{\perp}}{|q|B}
\label{eq:gyroradius}
\end{equation}

Where \( m \) is the particle's mass, and \( v_{\perp} \) is its velocity component perpendicular to \( \vec{B} \) \cite{chen1984introduction}.

One prominent configuration for achieving magnetic confinement is the tokamak. A toroidal (doughnut-shaped) magnetic field confines the plasma in this design. This magnetic field can be approximately described as Eq. \eqref{eq:tokamak_field}:

\begin{equation}
B_{\text{tokamak}} = B_{\text{external}} + \frac{\mu_0 I_p}{2\pi r}
\label{eq:tokamak_field}
\end{equation}

\( B_{\text{external}} \) represents the magnetic field from external coils, \( \mu_0 \) the permeability of free space, \( I_p \) the current passing through the plasma, and \( r \) the radial distance from the torus center \cite{wesson2011tokamaks}.

A significant parameter for assessing the effectiveness of magnetic confinement is the plasma beta, \( \beta \), defined as Eq. \eqref{eq:plasma_beta}:

\begin{equation}
\beta = \frac{n k T}{B^2/2\mu_0}
\label{eq:plasma_beta}
\end{equation}

Here, \( n \) denotes the plasma density, \( k \) the Boltzmann constant, and \( T \) the plasma temperature \cite{freidberg2007plasma}.

\subsection{Plasma Acceleration in MFPD}

In MFPDs, plasma is confined and accelerated to produce thrust. The fundamental principle behind this acceleration is the Lorentz force, as mentioned previously \cite{Chen1984}.

\subsubsection{Generation of Plasma in MFPD}

In the context of fusion propulsion, plasma generation is a prerequisite to initiate and sustain fusion reactions. For the MFPD, achieving the necessary conditions for nuclear fusion relies on effectively ionizing the fusion fuel. Here's how the MFPD system approaches plasma generation:

\begin{itemize}
    \item \textbf{Electrothermal Method:} Within the MFPD, an initial electrical discharge is passed through the fusion fuel, primarily deuterium and tritium, ionizing it. This method is employed to reach the preliminary ionization state before any confinement methods take over \cite{Nanayakkara2015}.
    
    \item \textbf{Magnetic Induction:} Post the initial ionization, changes in the magnetic fields within the MFPD can further induce currents in the plasma. This not only aids in maintaining a high level of ionization but also plays a crucial role in magnetic confinement, a core feature of the MFPD system \cite{Nanayakkara2015}.
\end{itemize}

It's imperative to note that while traditional electric propulsion (EP) systems and the MFPD both employ plasma generation techniques, their application and end goals differ. EP systems focus on propelling ionized propellant out of a thruster for propulsion, whereas the MFPD aims to achieve controlled fusion reactions by confining and sustaining a highly ionized plasma state \cite{Nanayakkara2015}.

\subsubsection{Acceleration Mechanism in MFPD}

The MFPD leverages the principle of the Lorentz force for plasma acceleration, but its specificity lies in how it configures the plasma and magnetic fields to achieve this effect.

\begin{itemize}
    \item \textbf{Plasma Geometry and Currents:} In the MFPD system, the plasma is shaped in a toroidal (donut-like) geometry. This configuration ensures a continuous and circulating plasma current, which is essential for sustaining fusion reactions. This circulating current, denoted as \( I_{plasma} \), interacts with the applied magnetic field, resulting in a Lorentz force that aids in both confinement and acceleration.

    \item \textbf{Magnetic Field Profile:} The magnetic field in the MFPD is a composite of the external field produced by superconducting magnets and the self-induced field due to the plasma current. The strength and direction of these fields are meticulously controlled to maximize the Lorentz force's effect, driving the plasma towards the magnetic nozzle and producing thrust.

    \item \textbf{Acceleration Using Lorentz Force:} The relationship between the plasma current density \( \mathbf{J} \), the magnetic field \( \mathbf{B} \), and the pressure gradient \( \nabla P \) can be articulated by Eq. \eqref{eq:acceleration}. In the MFPD, the term \( \mathbf{J} \times \mathbf{B} \) is maximized by optimizing the plasma geometry and magnetic field profile, ensuring efficient acceleration.
\end{itemize}

\begin{equation}
    \mathbf{J} \times \mathbf{B} = \nabla P
    \label{eq:acceleration}
\end{equation}

Where:

\begin{itemize}
    \item \( \mathbf{J} \) is the current density in the plasma.
    \item \( \mathbf{B} \) is the magnetic field.
    \item \( \nabla P \) represents the pressure gradient in the plasma.
\end{itemize}

The cross product \( \mathbf{J} \times \mathbf{B} \) yields the Lorentz force per unit volume acting on the plasma, which drives its acceleration, while \( \nabla P \) indicates the change in pressure across a certain distance in the plasma. The balance between these two terms is essential for efficient plasma confinement and acceleration.

The plasma's acceleration is thus a result of the strategic configuration and interplay between the plasma currents and the magnetic fields within the MFPD. This approach ensures not only effective propulsion but also aids in maintaining the required conditions for continuous fusion reactions.

\subsection{Thrust Generation Mechanism of MFPD}

The fundamental principle for any propulsion system in space relies on Newton's third law: to accelerate in one direction, a spacecraft must expel mass in the opposite direction. In the context of the MFPD, this expelled mass comes in the form of hot, charged particles – plasma. The mechanism that enables this expulsion is multi-faceted:

\subsubsection{Creation of High-Energy Plasma}

The initial step in thrust generation is to create a high-energy plasma. Fusion reactions provide the energy source for this plasma. When light atomic nuclei, typically isotopes of hydrogen-like deuterium and tritium, are fused under extreme conditions, they form helium and release a neutron and a significant amount of energy (Eq. \eqref{eq:fusion_reaction}. This reaction releases energy primarily in the form of kinetic energy of the produced helium and neutron \cite{glendinning2004high}. The helium ions (or alpha particles) are fully ionized and are confined within the magnetic field, thereby increasing the plasma's energy.

\subsubsection{Expelling the Plasma}

With the plasma heated to sufficient temperatures and pressures by fusion reactions, expelling it to produce thrust becomes essential. The plasma, being charged, responds strongly to magnetic and electric fields. By designing a suitable magnetic nozzle, the high-energy plasma can be directed and expelled from the spacecraft, producing thrust. 

The principle is analogous to a de Laval nozzle in a chemical rocket, where the shape of the nozzle accelerates the exhaust gases and directs them in a specific direction. However, the magnetic nozzle employs magnetic fields instead of physical walls to contain and direct the plasma. 

The thrust, \( T \), is given by:

\begin{equation}
T = \dot{m} v_e + (p_e - p_0) A_e
\label{eq:thrust_equation}
\end{equation}

Where \( \dot{m} \) is the mass flow rate of the plasma, \( v_e \) is the exhaust velocity, \( p_e \) and \( p_0 \) are the exhaust and ambient pressures respectively, and \( A_e \) is the nozzle exit area \cite{humble2005space}.

\subsubsection{Advantages of Fusion-Driven Propulsion}

The exhaust velocity \( v_e \) is a critical parameter as it determines the propulsion system's efficiency. The fusion-driven propulsion can achieve much higher exhaust velocities compared to chemical rockets. Higher \( v_e \) means that for a given amount of thrust, the spacecraft needs to expel less mass, making the propulsion system much more mass-efficient.

The specific impulse, \( I_{sp} \), a measure of propulsive efficiency, is related to the exhaust velocity (Eq. \eqref{eq:specific_impulse_2}):

\begin{equation}
I_{sp} = \frac{v_e}{g_0}
\label{eq:specific_impulse_2}
\end{equation}

Where \( g_0 \) is the standard acceleration due to gravity. Fusion-driven systems can achieve \( I_{sp} \) values several times greater than chemical propulsion systems \cite{frisbee2003advanced}.

\subsection{Mass Budget and Neutron Shielding in MFPD}

\subsubsection{Importance of Neutron Shielding}

The primary fusion reaction in the MFPD involves deuterium and tritium (Eq. \eqref{eq:fusion_reaction}). This reaction yields a high-energy neutron, as in equation \eqref{eq:fusion_reaction}. These neutrons are not confined by the magnetic fields (as they are neutral) and can penetrate deep into materials, causing structural damage, inducing radioactivity, and posing threats to human health \cite{mitchell2013radiation, slater2018neutron}.

\subsubsection{Shielding Materials}

The ideal shielding material for neutrons would possess the following characteristics:

\begin{itemize}
    \item High cross-section for neutron absorption.
    \item Capacity to slow down fast neutrons to thermal energies, where they can be easily absorbed.
    \item Low secondary gamma-ray production.
    \item Structural integrity under irradiation.
    \item Lightweight for space applications.
\end{itemize}

Hydrogenous materials, such as polyethylene, are effective at slowing down fast neutrons due to their similar mass to the neutron. Additionally, compounds like boron carbide (\( B_4C \)) can absorb neutrons with minimal secondary gamma production \cite{turner2013atoms}.

\subsubsection{Mass Budget Estimation}

The mass of the neutron shield is determined by the required attenuation factor and the material's properties. For instance, considering a desired attenuation factor \( AF \) (a reduction factor of the incoming neutron flux), the mass thickness \( m \) can be estimated by Eq. \eqref{eq:mass_thickness}:

\begin{equation}
m = \frac{-\ln(AF)}{\Sigma}
\label{eq:mass_thickness}
\end{equation}

Where \( \Sigma \) is the macroscopic cross-section of the shielding material.

However, neutron shielding isn't the only component contributing to the mass budget. The total mass budget \( M \) can be conceptualized as Eq. \eqref{eq:mass_budget}:

\begin{equation}
M = M_{\text{MFPD core}} + M_{\text{shielding}} + M_{\text{support structures}} + \ldots
\label{eq:mass_budget}
\end{equation}

Where:
\begin{itemize}
    \item \( M_{\text{MFPD core}} \) is the mass of the core propulsion system.
    \item \( M_{\text{shielding}} \) is the mass of the neutron and other radiation shields.
    \item \( M_{\text{support structures}} \) encompasses the mass of structural elements, conduits, coolant systems, and other auxiliary systems.
\end{itemize}

Note: The exact masses would depend on detailed design specifications, materials chosen, and engineering constraints.

\subsection{Thrust Generation Mechanism in MFPD}

\subsubsection{Fusion Reactions and High-Energy Particles}

The primary reaction driving the MFPD involves deuterium and tritium fusion (Eq. \eqref{eq:fusion_reaction}). The fusion of these nuclei in equation releases a neutron and a helium nucleus (or alpha particle) with significant kinetic energy. This energy, in the form of high-speed particles, is central to the thrust generation process.

\subsubsection{Harnessing Plasma Thrust}
The high-speed helium nuclei (alpha particles) produced from the fusion reactions serve as the primary propellant in the MFPD. They are charged particles, which means they can be manipulated using magnetic fields. The principle behind MFPD's thrust generation is to extract the kinetic energy of these particles and expel them at high velocities, generating thrust via Newton's third law \cite{chen1984introduction}.

A magnetic nozzle is utilized for this purpose. This nozzle converts the thermal and kinetic energy of the plasma into directed kinetic energy, expelling the plasma at high velocities and producing thrust (Eq. \eqref{eq:thrust_eqn}).

\begin{equation}
F_{\text{thrust}} = \dot{m} v_{\text{exhaust}} + (P_{\text{exit}} - P_{\text{ambient}})A_{\text{exit}}
\label{eq:thrust_eqn}
\end{equation}

Where:
\begin{itemize}
    \item \( F_{\text{thrust}} \) is the thrust.
    \item \( \dot{m} \) is the mass flow rate of the expelled plasma.
    \item \( v_{\text{exhaust}} \) is the exhaust velocity of the plasma.
    \item \( P_{\text{exit}} \) and \( P_{\text{ambient}} \) are the pressures at the nozzle exit and ambient space, respectively.
    \item \( A_{\text{exit}} \) is the area of the nozzle exit.
\end{itemize}

The primary thrust component arises from the high exhaust velocity of the plasma, while the pressure differential term becomes significant only in low vacuum environments and negligible in deep space.

\subsubsection{Magnetic Nozzle and Plasma Expansion}

A magnetic nozzle doesn't have solid walls like traditional nozzles. Instead, it uses magnetic fields to guide and accelerate the plasma. As the plasma moves through the diverging magnetic field, it expands, and its particles are accelerated due to the conservation of magnetic moment, leading to an increase in the particle's perpendicular kinetic energy and a conversion of this energy into directed flow energy \cite{ahedo2004plasma}. This magnetic guidance ensures that the plasma is expelled, controlled, and directed, maximizing thrust and preventing plasma losses to the spacecraft's structure.

\subsection{Design Considerations for MFPD Thrusters}
Magnetic Fusion Plasma Drive (MFPD) thrusters promise a new frontier in propulsion capabilities, but their implementation and efficiency hinge on careful design and meticulous engineering. With the potential to revolutionize propulsion technology, the MFPD also brings with it challenges that demand innovative solutions. This section outlines key design considerations for MFPD thrusters, highlighting the challenges and prospective solutions in ensuring these propulsion systems' optimal performance, longevity, and energy efficiency.

\subsubsection{Role of Electrodes in MFPD}

In MFPD systems, electrodes play a crucial role in initiating and maintaining the plasma. They serve as a medium to inject current into the plasma, which can either help in ionizing the propellant or in maintaining the fusion process, depending on the design specifics of the propulsion system.

\begin{itemize}
    \item \textbf{Location and Configuration}: Electrodes are strategically placed in the thruster chamber, often at the entrance or close to the propellant injection site. Their placement ensures optimal interaction with the incoming propellant and existing plasma. 

    \item \textbf{Operating Envelope}: These electrodes operate under extreme conditions with temperatures exceeding thousands of Kelvin and exposure to high-energy plasma particles. As a result, they are subject to wear and erosion over time.

    \item \textbf{Purpose}: Their primary purpose is to introduce an electric current into the propellant. This current, in conjunction with applied or inherent magnetic fields, helps in the ionization of the propellant, sustenance of the fusion process, and acceleration of the plasma for thrust generation.
\end{itemize}

Given their critical role, electrode erosion emerges as a significant challenge in MFPD systems.

\subsubsection{Electrode Erosion}

Continuous exposure to high-energy plasma leads to the degradation of electrodes over time. This erosion not only affects the longevity of the MFPD system but can also introduce impurities into the plasma, impacting its performance. Addressing this challenge requires:

\begin{itemize}
    \item \textbf{Cooling Systems}: Implementing active cooling mechanisms for the electrodes can significantly prolong their operational lifespan by reducing thermal stresses and sputtering effects \cite{Sankaran2014}.
    
    \item \textbf{Material Selection}: Opting for materials with higher melting points and lower sputtering yields can ensure that electrodes retain their structural integrity for longer periods. Materials such as tungsten or graphite, commonly used in other plasma-facing components, might be suitable choices \cite{Sankaran2014}.
    
        \item \textbf{Electrodeless Designs in Fusion Propulsion}: While the specific design of the MFPD we are discussing utilizes electrodes, it's worth noting the broader landscape of fusion propulsion technologies. In the domain of fusion propulsion, some approaches are being explored that do not rely on solid electrodes to initiate and sustain plasma. These designs aim to address the erosion issue inherent to systems with electrodes. While commonly associated with Electric Propulsion (EP) systems, methods such as radio-frequency or microwave ionization have also been researched for potential application in fusion propulsion. However, the implementation in fusion systems presents its own set of challenges and is outside the primary focus of our discussion on the MFPD system \cite{Sankaran2014}. In this paper, MFPD refers to a specific novel fusion propulsion system that uses superconducting magnets for plasma confinement and magnetic nozzles for thrust generation rather than a generic label for any fusion propulsion system.

\end{itemize}

\subsubsection{Power Supply}

MFPD thrusters require significant power. The source of this power can be:

\begin{itemize}
    \item Solar Arrays: Suitable for low to moderate power requirements \cite{Choina2020}.
    
    \item Nuclear Reactors: For high-power applications, especially in deep space where sunlight is scarce \cite{Choina2020}.
\end{itemize}

\subsubsection{Magnetic Field Generation and Fusion Power in the MFPD}

\subsubsection{Magnetic Field Generation and Fusion Power in the MFPD}

In the MFPD system, fusion reactions serve dual purposes: producing high-energy plasma for propulsion and generating electricity for onboard systems, including magnetic field generation. The fusion process yields high-energy plasma that can be expelled for thrust and produces neutrons which can be captured in a blanket, triggering further reactions that release heat. This heat can be converted to electricity using thermoelectric generators or other methods. Herein lies the distinction of the MFPD from typical EP systems.

\begin{itemize}
    \item \textbf{Self-Sustaining Power Generation}: Once the fusion reactions are initiated, the MFPD uses the energy produced by the fusion process itself to generate the electricity required to sustain the magnetic fields. This potential closed-loop system contrasts with traditional EP systems that rely on external power sources, such as solar panels or nuclear reactors. Fusion's inherent energy density enables this self-sustenance, a characteristic absent in traditional EP systems.
    
    \item \textbf{High Thrust and Efficiency}: Fusion reactions, given their immense energy release, allow the MFPD to potentially provide both high specific impulse and high thrust, a combination challenging to achieve with existing propulsion systems.
    
    \item \textbf{Dual Utility}: The MFPD is a propulsion system and power generator for spacecraft subsystems, reducing the need for additional onboard power generation methods.
    
    \item \textbf{Choice of Magnetic Field Generation}: Depending on mission requirements, the MFPD can leverage self-generated magnetic fields, a concept challenging to achieve but under active investigation in the fusion community. These fields can be augmented with externally applied fields using electricity derived from fusion reactions for more demanding operations.
\end{itemize}

It's crucial to emphasize that while the conceptual benefits of MFPD are significant, the technology is still nascent, especially when compared to more mature EP systems. However, its potential merits warrant further research and development.

\subsection{Design of the Magnetic Fusion Plasma Drive (MFPD)}
\begin{figure}[h]
    \centering
    \includegraphics[width=0.8\textwidth]{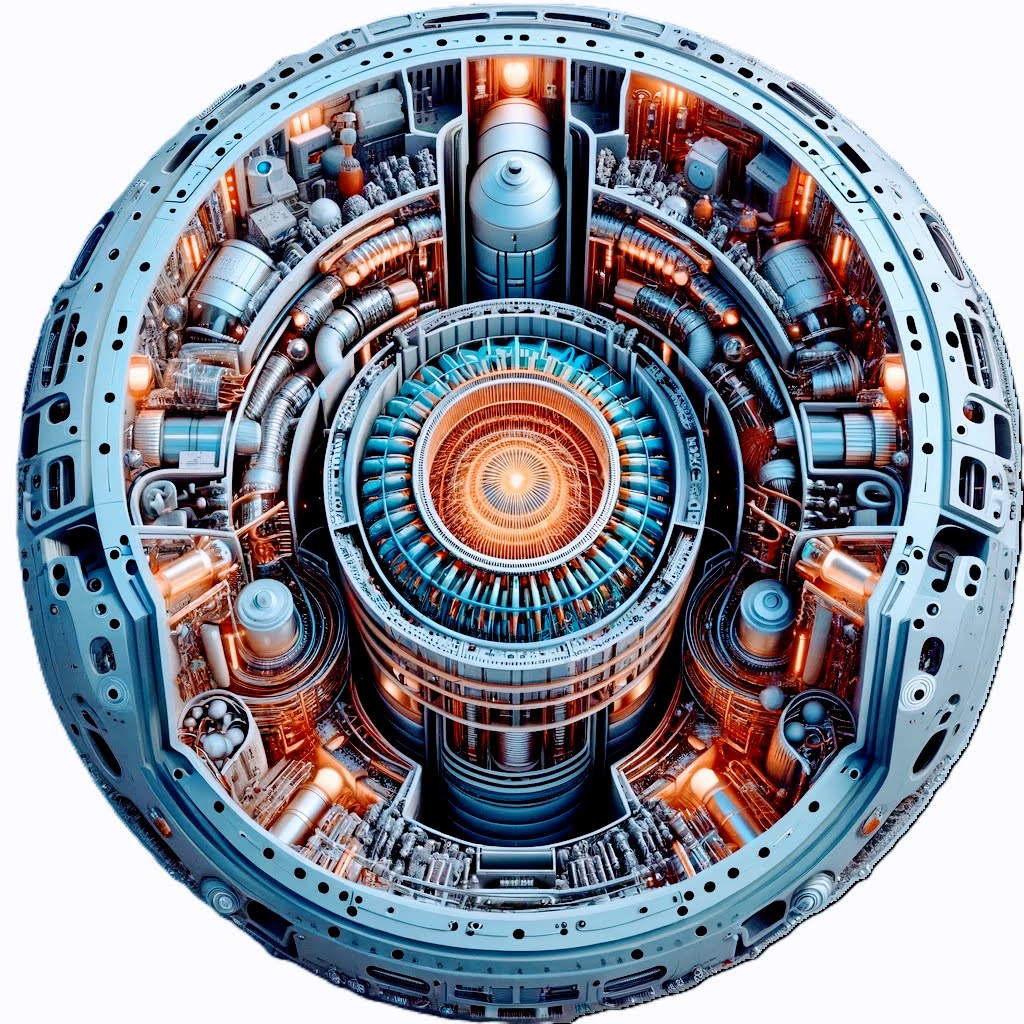}
    \caption{A 3D cutaway illustration of the MFPD, as seen from the front.}
    \label{fig:mfpd}
\end{figure}

The Magnetic Fusion Plasma Drive (MFPD) represents a significant advancement in space propulsion technology, leveraging the principles of nuclear fusion. Fig. \ref{fig:mfpd} illustrate the conceptual design of the MFPD, highlighting its key components and operational mechanisms.

\begin{itemize}
    \item \textbf{Toroidal Fusion Reactor Core:} The core of the MFPD is a toroidally shaped fusion reactor where deuterium and tritium are fused. This core is responsible for initiating and sustaining the fusion reactions, with visible plasma indicating the high-energy environment necessary for fusion. The core contains visible plasma in Fig. \ref{fig:mfpd}, indicating the high-energy state necessary for fusion reactions.
    
    \item \textbf{Fuel Injection System:} Deuterium and tritium, the primary fusion fuels, are injected into the reactor core through a specialized fuel injection system. This system ensures the precise and controlled fuel delivery to the fusion site.
    
    \item \textbf{Magnetic Confinement System:} Surrounding the reactor core are superconducting magnets, vital for creating the magnetic fields necessary for plasma confinement. These fields contain and stabilize the plasma, maintaining it in a toroidal geometry essential for efficient fusion. In Fig. \ref{fig:mfpd}, surrounding the reactor core, superconducting magnets are depicted.
    
    \item \textbf{Plasma Heating and Ignition Systems:} The MFPD employs electromagnetic induction or radio frequency heating methods to achieve the temperatures required for initiating fusion reactions. These systems are crucial in elevating the plasma to the required energy state. In Fig. \ref{fig:mfpd}, systems like electromagnetic induction or radio frequency heating, which are crucial for achieving the high temperatures needed for fusion, are shown.
    
    \item \textbf{Magnetic Nozzle:} A key feature of the MFPD is its magnetic nozzle, which guides and accelerates the plasma without needing physical walls. This nozzle converts the plasma's thermal and kinetic energy into directed flow energy, which is critical for propulsion. In Fig. \ref{fig:mfpd}, the magnetic nozzle is not explicitly visualized but would be situated at the exit point of the fusion reactor, where the plasma is expelled. 
    
    \item \textbf{Neutron Shielding:} Given the production of high-energy neutrons during fusion, the MFPD incorporates neutron shielding, made from materials like polyethylene and boron carbide. This shielding is crucial for protecting the spacecraft and its occupants. In Fig. \ref{fig:mfpd}, neutron shielding is illustrated, made from materials like polyethylene and boron carbide, to protect the spacecraft and its occupants from high-energy neutrons produced during fusion.
    
    \item \textbf{Support Structures:} The engine includes various support structures, such as conduits and cooling systems, which are integral to the operation and maintenance of the MFPD.
    
    \item \textbf{Power Generation Systems:} While not explicitly depicted in the illustration, the MFPD design accommodates power generation units, which may be solar arrays or nuclear reactors, depending on mission-specific requirements.
\end{itemize}

The integration of these components exemplifies the sophisticated engineering and design considerations inherent in the development of an advanced space propulsion system like the MFPD.

\subsection{Future of MFPD Thrusters}

With advances in power generation and materials science, MFPD thrusters are poised to play a significant role in future space missions. Their ability to provide high thrust combined with good efficiency makes them attractive for various mission profiles, from satellite station-keeping to deep space exploration.

The MFPD offers a promising avenue for electric propulsion, leveraging electromagnetic principles to accelerate plasma and produce thrust. While challenges remain in electrode erosion and power requirements, ongoing research and technological advancements hint at a bright future for MFPD propulsion in space exploration.

\section{Mathematical Formulations of the MFPD System}
The MFPD system, with its potential to redefine space propulsion, operates on intricate principles grounded in physics and mathematics. As with any novel technology, a rigorous mathematical treatment is imperative to comprehend and fine-tune its performance. This section systematically breaks down the core mathematical relationships governing the MFPD system, ranging from the foundational equations of fusion reactions and plasma thrust to the intricate interplay of magnetic fields and plasma confinement. By dissecting these formulations, we aim to establish a theoretical framework that not only elucidates the mechanics of the MFPD system but also lays the groundwork for its optimization and further innovations.

\subsection{Fusion Reaction Rates}

Quantum mechanics dictates fusion reactions, but for macroscopic rates, we often use cross-sections averaged over thermal distributions of particle speeds. This leads to the concept of reactivity (Eq. \eqref{eq:reactivity}):

\begin{equation}
    \langle \sigma v \rangle = \int_0^{\infty} \sigma(v) v f(v) dv
    \label{eq:reactivity}
\end{equation}

Where \( f(v) \) represents the Maxwellian velocity distribution (Eq. \eqref{eq:maxwellian}) \cite{Chen1984}:

\begin{equation}
    f(v) = \left( \frac{m}{2 \pi k_B T} \right)^{3/2} 4\pi v^2 e^{-\frac{mv^2}{2k_BT}}
    \label{eq:maxwellian}
\end{equation}

Given this, the fusion power density \( P \) can be expressed as Eq. \eqref{eq:power_density} \cite{Chen1984}:

\begin{equation}
    P = \frac{1}{2} n^2 \langle \sigma v \rangle E_{fusion}
    \label{eq:power_density}
\end{equation}

Where \( E_{fusion} \) is the energy released per fusion reaction.

The efficiency of magnetic confinement directly impacts the fusion rate. Particle confinement time \( \tau \) measures how long, on average, a particle remains in the plasma before it's lost. It's a crucial parameter, and the product \( n\tau \) often serves as a benchmark for the viability of sustained fusion \cite{Hutchinson2008}.

\subsection{Plasma Thrust Equations}

The fundamental equations governing the behavior of a plasma in an electromagnetic field are the fluid equations coupled with Maxwell's equations. For a quasi-neutral plasma with inertial effects neglected, the fluid momentum equation becomes Eq. \eqref{eq:momentum} \cite{Chen1984}:

\begin{equation}
    m_i n \left( \frac{\partial \mathbf{V}}{\partial t} + \mathbf{V} \cdot \nabla \mathbf{V} \right) = -n \nabla U + q_n (\mathbf{E} + \mathbf{V} \times \mathbf{B}) - \nabla \cdot \mathbf{P}
    \label{eq:momentum}
\end{equation}

Here, \( m_i \) is ion mass, \( \mathbf{V} \) is plasma velocity, \( U \) is potential energy, \( q_n \) is charge number, \( \mathbf{E} \) and \( \mathbf{B} \) are electric and magnetic fields, and \( \mathbf{P} \) is the pressure tensor.

Using this, the change in momentum \( \Delta p \) produced by the MFPD, as a function of the fusion energy and the propellant mass, is given by:

\begin{equation}
\Delta p = \sqrt{2 \times m_{\text{propellant}} \times E_{\text{fusion}}} 
\label{eq:momentum_change}
\end{equation}

This momentum change, in combination with the nozzle design and propellant flow rates, gives rise to the thrust \( T \) produced by the MFPD:

\begin{equation}
    T = \dot{m} V_{exit} = A_p P_{exit}
    \label{eq:thrust2}
\end{equation}

Where \( A_p \) is the area of the propulsion nozzle and \( P_{exit} \) is the plasma pressure at the nozzle exit \cite{Chen1984}.

For optimal performance, the specific impulse \( I_{sp} \) is defined as Eq. \eqref{eq:isp} \cite{Chen1984}:

\begin{equation}
    I_{sp} = \frac{V_{exit}}{g_0}
    \label{eq:isp}
\end{equation}

Here, \( g_0 \) is the gravitational acceleration constant.

\subsection{Magnetic Field Equations}

The confinement and manipulation of plasma in an MFPD system are intrinsically tied to the properties of the magnetic fields employed. Maxwell's equations form the foundation for the behavior of these magnetic fields in the presence of currents and charges (Eq. \eqref{eq:divergence_b}):

\begin{equation}
    \nabla \cdot \mathbf{B} = 0
    \label{eq:divergence_b}
\end{equation}

This equation asserts no magnetic monopoles exist; the magnetic field lines are continuous and closed (Eq. \eqref{eq:curl_b}) \cite{Griffiths2017}.

\begin{equation}
    \nabla \times \mathbf{B} = \mu_0 \mathbf{J} + \mu_0 \epsilon_0 \frac{\partial \mathbf{E}}{\partial t}
    \label{eq:curl_b}
\end{equation}

Here, \( \mu_0 \) is the permeability of free space, \( \mathbf{J} \) is the current density, \( \epsilon_0 \) is the permittivity of free space, and \( \mathbf{E} \) is the electric field. This equation, Ampère's law with Maxwell's addition, relates the magnetic field to currents and changing electric fields \cite{Griffiths2017}.

For an MFPD system, magnetic confinement can be described using the safety factor, \( q \), which is crucial for assessing plasma stability (Eq. \eqref{eq:safety_factor}):

\begin{equation}
    q(a) = \frac{aB_t}{RB_p}
    \label{eq:safety_factor}
\end{equation}

Where \( R \) is the major radius of the torus and \( B_t \) and \( B_p \) are the toroidal and poloidal magnetic field components, respectively. Understanding the value of \( q \) across the plasma profile helps predict potential instabilities, with specific values and profiles preferred for stability \cite{Hutchinson2008}. Magnetic confinement is often modeled using the Grad-Shafranov equation, which describes equilibria in magnetically confined plasmas (Eq. \eqref{eq:grad_shafranov}) \cite{Freidberg2014}:

\begin{equation}
    \Delta^* \Psi = -\mu_0 R^2 \frac{dp}{d\Psi} - F \frac{dF}{d\Psi}
    \label{eq:grad_shafranov}
\end{equation}

Here, \( \Psi \) is the poloidal magnetic flux, \( p \) is the plasma pressure, \( R \) is the major radius, and \( F \) is a function representing the toroidal current inside a magnetic surface. The challenge with magnetic confinement is not just to confine the plasma but to do so stably over long periods. This often requires superimposing additional magnetic fields, analyzing various MHD stability modes, and even using feedback control mechanisms \cite{Hutchinson2008}.

This section provided a mathematical overview of the key aspects underlying the operation of a Magnetic Plasma Drive system. Understanding these equations is pivotal to model, predict, and control the behavior of such propulsion systems. A comprehensive grasp of these principles will be instrumental in advancing and optimizing MFPD technologies.

\section{Comparison of MFPD with Existing Fusion Propulsion Concepts}

Several key concepts have been proposed over the years in the realm of fusion-based propulsion. Our Magnetif Fusion Plasma Drive (MFPD) shares some similarities but also presents unique characteristics. Here, we delineate the primary distinctions between MFPD and other notable fusion propulsion concepts.

\subsection{Bussard Ramjet}
\begin{itemize}
    \item Concept: Proposed by Robert W. Bussard in 1960, this interstellar fusion rocket collects hydrogen from space with a magnetic ramscoop to use as fusion fuel \cite{Bussard1960}.
    \item Comparison: Unlike the MFPD, the Bussard Ramjet relies on collecting its fuel from the interstellar medium. While the Ramjet is designed for interstellar distances, MFPD is envisioned primarily for interplanetary missions.
\end{itemize}

\subsection{Direct Fusion Drive (DFD)}

\begin{itemize}
    \item Concept: Developed primarily by Princeton Satellite Systems, the Direct Fusion Drive (DFD) is a propulsion system that simultaneously provides thrust and electric power. It uses a fusion reaction involving deuterium and other isotopes \cite{Cohen2013}. 
    Field-Reversed Configuration (FRC) is a method to confine plasma without needing an external magnetic field, using the plasma's self-induced magnetic field. In DFD, a field-reversed configuration is employed to achieve compact and efficient confinement of the fusion fuel. The thrust in DFD is primarily generated by expelling deuterium that has been heated by the hot plasma surrounding the fusion reaction. Due to its high temperature, this plasma is expelled at high velocities, generating a thrust as per Newton's third law.
    
    \item Comparison with MFPD: DFD and MFPD employ magnetic confinement as central components. However, they differ in their fusion reactions and confinement strategies. While DFD relies on deuterium and adopts a field-reversed configuration for confinement, the MFPD focuses on D-T (Deuterium-Tritium) fusion and utilizes a generalized magnetic confinement approach.
\end{itemize}

\subsection{Magnetic Target Fusion (MTF)}
\begin{itemize}
    \item Concept: MTF employs magnetic fields to compress fusion fuel, followed by ignition through lasers or other external agents \cite{Lindemuth2000}.
    \item Comparison: MFPD primarily focuses on harnessing magnetic fields for controlling plasma for propulsion, not necessarily for fusion ignition, marking a departure from the MTF methodology.
\end{itemize}

\subsection{VASIMR}
\begin{itemize}
    \item Concept: A creation of the Ad Astra Rocket Company, VASIMR utilizes radio waves to ionize propellant, with magnetic fields subsequently accelerating the plasma to generate thrust \cite{ChangDiaz2000}.
    \item Comparison: While both models employ plasma and magnetic fields, VASIMR's thrust generation does not hinge on fusion reactions, setting it apart from the MFPD.
\end{itemize}

\subsection{Project Daedalus}
\begin{itemize}
    \item Concept: A brainchild of the British Interplanetary Society from the 1970s, Daedalus was a proposed uncrewed interstellar probe utilizing fusion for propulsion, specifically deuterium/helium-3 fusion \cite{Bond1978}.
    \item Comparison: While both concepts leverage fusion for propulsion, Daedalus had interstellar travel in its crosshairs and employed D/He-3 fusion. In contrast, MFPD targets interplanetary missions with a broader D-T fusion perspective.
\end{itemize}

Based on the theoretical foundations and the principles underlying the MFPD system, certain potential advantages emerge in the context of fusion propulsion. As described, one distinguishing feature of the MFPD system is the proposed use of superconducting magnets for plasma confinement and magnetic nozzles for thrust generation. While these features might lead to enhanced thrust capabilities and potentially reduced fusion fuel requirements, these are based on initial analyses and require further in-depth study and simulation for validation. Like other fusion propulsion methods, the MFPD system is expected to face challenges, especially in areas such as controlled fusion in the space environment and materials science concerns. However, if realized, the potential for higher specific impulses could position the MFPD system and fusion propulsion at large as attractive candidates for prolonged space missions. It's crucial to note that the advantages and challenges highlighted here are based on preliminary assessments, and comprehensive simulations and analyses are needed to ascertain these claims conclusively.

\section{Example Calculations and Descriptions}

\subsection{MFPD Plasma Dynamics}

The behavior of plasma within the Magnetoplasmadynamic (MFPD) propulsion system can be primarily understood through magnetohydrodynamics (MHD). In the MFPD, the presence of an externally applied magnetic field and electric currents through the plasma results in a Lorentz force which accelerates the plasma out of the thruster.

Using the MHD momentum equation:
\begin{equation}
\rho \left( \frac{\partial \mathbf{V}}{\partial t} + \mathbf{V} \cdot \nabla \mathbf{V} \right) = -\nabla P + \mathbf{J} \times \mathbf{B} - \nabla \cdot \mathbf{\Pi}
\end{equation}
where \( \rho \) is the plasma density, \( \mathbf{V} \) is the plasma velocity, \( P \) is the pressure, \( \mathbf{J} \) is the current density, \( \mathbf{B} \) is the magnetic field, and \( \mathbf{\Pi} \) is the viscous stress tensor.

In the context of the MFPD, the dominant term is the Lorentz force, \( \mathbf{J} \times \mathbf{B} \). This force accelerates the plasma, providing thrust to the spacecraft. The electric currents that give rise to \( \mathbf{J} \) are induced by the fusion reactions taking place within the MFPD chamber, making it imperative to maintain conditions conducive to fusion.

To maintain a quasi-neutral plasma and to aid in achieving conditions for fusion, the magnetic confinement becomes crucial. It not only ensures that the charged plasma particles remain within the thruster chamber long enough for fusion reactions but also drives the particles out at high velocities due to the Lorentz force. The magnetic confinement time and the particle densities dictate the efficiency of fusion reactions and, subsequently, the efficiency of the MFPD.

It's worth noting that the efficiency of the MFPD system is intrinsically linked to the balance between the magnetic confinement, ensuring sufficient fusion reactions, and the Lorentz force, which provides the thrust. Any variations in the plasma parameters, such as density, temperature, or electric conductivity, will directly influence the MFPD's thrust and efficiency.

In summary, the plasma dynamics within the MFPD thruster revolve around harnessing the Lorentz force to achieve efficient propulsion. This force is the culmination of fusion-driven electric currents and the externally applied magnetic field. Understanding and optimizing these dynamics are pivotal for realizing the potential advantages of the MFPD system for space missions.

\subsection{Fusion Processes in MFPD}

The Magnetoplasmadynamic (MFPD) propulsion system capitalizes on the energy released from fusion reactions to generate thrust. Here, we delve into the specifics of these fusion reactions and the conditions required for their occurrence within the MFPD.

\subsubsection{Deuterium-Tritium Fusion}

As noted, our focus is on deuterium-tritium (D-T) fusion, which is one of the most energetically favorable fusion reactions. The reaction is represented as:

\begin{equation}
D + T \rightarrow \alpha + n + 17.6 \, \text{MeV}
\end{equation}

Here, \( \alpha \) is a helium nucleus, and \( n \) is a neutron. The energy of 17.6 MeV is distributed among the products, with the majority carried away by the neutron.

\subsubsection{Conditions for Fusion}

For fusion to occur, the plasma within the MFPD must attain conditions referred to as the Lawson criterion. This involves achieving a critical product of plasma density and confinement time. The equation representing the Lawson criterion is:

\begin{equation}
n \tau \geq \frac{12kT}{E_{\text{fusion}}}
\end{equation}

Where:
- \( n \) is the number density of the fusion fuel.
- \( \tau \) is the energy confinement time.
- \( k \) is the Boltzmann constant.
- \( T \) is the plasma temperature.
- \( E_{\text{fusion}} \) is the energy released per fusion reaction.

Given the high temperatures required for D-T fusion, the fusion fuels are fully ionized, forming a plasma of positively charged nuclei and free electrons. At these temperatures, the coulombic repulsion between the positively charged D and T nuclei becomes significant. Thus, the plasma needs to be sufficiently hot, dense, and confined for a long enough time to overcome this repulsion and allow fusion to take place.

\subsubsection{Challenges and Constraints}

Fusion within the MFPD faces challenges that include:

\begin{itemize}
    \item Magnetic Confinement: Achieving a strong and stable magnetic field that can confine the high-temperature plasma efficiently.
    \item Radiative Losses: High-temperature plasmas emit radiation, leading to energy losses that can inhibit fusion reactions.

    \item Fuel Recycling: Capturing and recycling unburnt fusion fuel to maximize fuel efficiency.

    \item Neutron Management: The neutrons produced in D-T fusion are not confined by magnetic fields and can escape, leading to radiation hazards and potential structural damage to the MFPD.
\end{itemize}

\subsection{Performance Metrics}

We use standardized performance metrics to objectively compare the MFPD propulsion system against conventional chemical rockets. These metrics give insights into the efficiency, capability, and potential advantages of the MFPD system for long-duration space missions.

\subsubsection{Specific Impulse (\(I_{sp}\))}

Specific Impulse is a measure of the efficiency of a propulsion system. It is defined as the total impulse delivered per unit weight of propellant consumed:

\begin{equation}
I_{sp} = \frac{\Delta v}{g_0 \cdot \ln\left(\frac{m_0}{m_f}\right)}
\end{equation}

Where:
- \( \Delta v \) is the change in velocity.
- \( g_0 \) is the standard gravitational acceleration (approximately \(9.81 \, \text{m/s}^2\)).
- \( m_0 \) is the initial mass of the spacecraft.
- \( m_f \) is the final mass of the spacecraft after propellant consumption.

Higher \(I_{sp}\) values represent better fuel efficiency. The MFPD system's ability to attain significantly higher \(I_{sp}\) values than chemical rockets makes it particularly suitable for long-duration missions.

\subsubsection{Thrust-to-Weight Ratio}

The thrust-to-weight ratio is a dimensionless parameter indicating the propulsion system's performance concerning its weight. For spacecraft propulsion, especially for interstellar missions, a higher thrust-to-weight ratio can be crucial:

\begin{equation}
\frac{T}{W} = \frac{\text{Thrust produced by the engine}}{\text{Weight of the propulsion system}}
\end{equation}

\subsubsection{Fuel Efficiency}

While \(I_{sp}\) provides insights into the efficiency concerning propellant consumption, fuel efficiency looks at the energy extracted from the fuel relative to the total energy available in that fuel. For fusion reactions, this metric becomes critical given the high-energy yields of fusion fuels.

\subsubsection{Endurance}

Endurance, in this context, refers to the ability of the propulsion system to sustain thrust over extended periods. Given that space missions can last months to years, the longevity of the propulsion system without significant degradation is pivotal.

\subsubsection{Operational Flexibility}

This metric considers the system's ability to adapt to different mission profiles. It examines aspects like throttleability, start-stop cycles, and adaptability to different power levels.

\subsubsection{Safety and Reliability}

Especially crucial for manned missions, this metric assesses the risks associated with the propulsion system, including radiation hazards, potential system failures, and challenges in emergency shutdowns.

\subsubsection{Payload Fraction}

Given by the ratio of the payload mass to the total spacecraft mass, a higher payload fraction indicates a greater proportion of the spacecraft's mass is dedicated to the actual mission (instruments, crew, supplies), rather than propulsion or support systems.

\begin{equation}
\text{Payload Fraction} = \frac{m_{\text{payload}}}{m_0}
\end{equation}

Where \( m_{\text{payload}} \) is the mass of the payload.

In conclusion, by evaluating the MFPD propulsion system through these standardized metrics, we can understand its advantages, potential limitations, and areas for optimization. This comprehensive analysis facilitates informed decisions about the suitability of the MFPD system for specific mission profiles and objectives.

\subsubsection{MFPD's Fusion Strategy}

To optimize fusion reactions, the MFPD system uses a combination of magnetic confinement and inertial confinement. Magnetic fields confine the plasma and reduce losses due to transport phenomena, while the inertia of the fuel itself (aided by rapid heating and compression) ensures a high local density, promoting fusion reactions. The combination aims to create a "sweet spot" where fusion conditions are achieved and maintained for efficient propulsion.

In conclusion, harnessing fusion energy in the MFPD requires a careful balance between plasma confinement, fuel density, and temperature. Achieving and maintaining this balance is pivotal for the propulsion system's efficient operation and realizing its potential benefits for long-duration space missions.

\subsection{Implications for Mars Mission}

Utilizing an MFPD propulsion system for a mission to Mars brings forth several significant advantages over traditional chemical rockets, primarily due to the increased efficiency and higher thrust capability. To understand these advantages better, we can perform some example calculations using a spacecraft with a total mass of \( m_0 = 100 \) metric tonnes, of which \( m_{\text{payload}} = 20 \) metric tonnes is payload.

\subsubsection{Delta-V Requirements}

The Delta-V (\( \Delta v \)) requirements for a Mars mission can vary based on the mission profile, orbital dynamics, and propulsion system. For a typical transfer orbit from Earth to Mars, a \( \Delta v \) of approximately \( 4.3 \, \text{km/s} \) is required. 

\subsubsection{Propellant Mass Fraction (PMF)}

Using the rocket equation, we can determine the propellant mass fraction (\( \text{PMF} \)) required to achieve the necessary \( \Delta v \):

\begin{equation}
\Delta v = I_{sp} \cdot g_0 \cdot \ln\left(\frac{m_0}{m_f}\right)
\end{equation}

Given that MFPD has a significantly higher \( I_{sp} \) than chemical rockets, for the sake of illustration, let's assume an \( I_{sp} \) of \( 5000 \, \text{s} \) for MFPD. Rearranging the above equation, we find:

\begin{equation}
m_f = m_0 \cdot \exp\left(-\frac{\Delta v}{I_{sp} \cdot g_0}\right)
\end{equation}

Inserting the values:

\begin{equation}
m_f = 100 \times \exp\left(-\frac{4.3 \times 10^3}{5000 \times 9.81}\right) \approx 77.1 \, \text{tonnes}
\end{equation}

Thus, the mass of propellant required \( m_p \) is:

\begin{equation}
m_p = m_0 - m_f \approx 22.9 \, \text{tonnes}
\end{equation}

\subsubsection{Duration to Reach Mars with Different Accelerations}

For continuous thrust propulsion systems like MFPD, the time to reach Mars can be estimated by considering the spacecraft's average acceleration. Assuming Mars is at an average distance of \( 225 \times 10^6 \, \text{km} \) from Earth, we can calculate the transit times for different accelerations, taking into account that the spacecraft spends half the journey accelerating and the other half decelerating:

1. \(0.5g\) (\(4.905 \, \text{m/s}^2\)):
\begin{equation}
t_{0.5g} = 2 \times \sqrt{\frac{225 \times 10^9 \, \text{m}}{4.905}} \approx 303,000 \, \text{s} \approx 3.5 \, \text{days}
\end{equation}

2. \(1g\) (\(9.81 \, \text{m/s}^2\)):
\begin{equation}
t_{1g} = 2 \times \sqrt{\frac{225 \times 10^9 \, \text{m}}{9.81}} \approx 214,000 \, \text{s} \approx 2.5 \, \text{days}
\end{equation}

3. \(1.5g\) (\(14.715 \, \text{m/s}^2\)):
\begin{equation}
t_{1.5g} = 2 \times \sqrt{\frac{225 \times 10^9 \, \text{m}}{14.715}} \approx 175,000 \, \text{s} \approx 2 \, \text{days}
\end{equation}

4. \(2g\) (\(19.62 \, \text{m/s}^2\)):
\begin{equation}
t_{2g} = 2 \times \sqrt{\frac{225 \times 10^9 \, \text{m}}{19.62}} \approx 151,000 \, \text{s} \approx 1.75 \, \text{days}
\end{equation}

These calculations indicate that MFPD could significantly reduce the transit time to Mars compared to traditional chemical rockets, potentially taking just a few days depending on the level of sustained acceleration.

\subsubsection{Dynamic Acceleration Scheme}

Given the need to minimize travel time and respect human physiological limits, we propose a dynamic acceleration scheme consisting of multiple phases:

\begin{enumerate}
    \item Initial high-acceleration phase at 3g (for a short duration).
    \item Sustained acceleration phase at 1g (for the majority of the first half of the journey).
    \item Coasting phase at zero acceleration (if required, depending on the achieved velocity).
    \item Deceleration phase mirroring the acceleration phase (first at 1g, then increasing to 3g).
\end{enumerate}

\textbf{Phase 1: High Acceleration (3g)}

\begin{equation}
    a_{\text{high}} = 3 \times 9.81 \, \text{m/s}^2 = 29.43 \, \text{m/s}^2
\end{equation}
\begin{equation}
    t_{\text{high}} = 600 \, \text{s} \, (\text{10 minutes})
\end{equation}
\begin{equation}
    d_{\text{high}} = \frac{1}{2} a_{\text{high}} t_{\text{high}}^2
\end{equation}
\begin{equation}
    v_{\text{high}} = a_{\text{high}} t_{\text{high}}
\end{equation}

\textbf{Phase 2: Sustained Acceleration (1g)}

\begin{equation}
    D_{\text{total}} = \frac{225,000,000 \times 1000}{2} \, \text{m}
\end{equation}
\begin{equation}
    D_{\text{remaining}} = D_{\text{total}} - d_{\text{high}}
\end{equation}
\begin{equation}
    a_{\text{sustained}} = 9.81 \, \text{m/s}^2
\end{equation}
\begin{equation}
    t_{\text{sustained}} = \sqrt{\frac{2 \times D_{\text{remaining}}}{a_{\text{sustained}}}}
\end{equation}
\begin{equation}
    v_{\text{sustained}} = a_{\text{sustained}} \times t_{\text{sustained}}
\end{equation}

\textbf{Phase 3: Coasting Phase (if applicable)}

\begin{equation}
    v_{\text{total}} = v_{\text{high}} + v_{\text{sustained}}
\end{equation}
Determine coasting duration based on \( v_{\text{total}} \) and remaining distance.

\textbf{Phase 4: Deceleration Phase}

Mirrors the acceleration phase in reverse order, ensuring the total distance covered equals half the distance to Mars.

\begin{equation}
    t_{\text{total}} = 2 \times (t_{\text{high}} + t_{\text{sustained}} + t_{\text{coasting}})
\end{equation}

This dynamic acceleration scheme aims to achieve the fastest possible travel time to Mars while keeping accelerations within a range tolerable for humans. The exact values for each phase need to be iteratively calculated based on real-time mission data and spacecraft capabilities.

\subsubsection{Advantages and Challenges}

\begin{itemize}
    \item Shorter Transit Times: As demonstrated, MFPD offers significantly reduced travel times at higher sustained accelerations, minimizing crew exposure to space radiation and reducing mission risks.
    \item Higher Payload Capacities: The efficiency of MFPD allows a larger fraction of the spacecraft's mass to be dedicated to payload, enhancing mission capabilities.
    \item Flexibility: MFPD systems can be throttled, allowing for adaptive mission profiles and potential abort scenarios.
    \item Challenges: High energy requirements and the necessity for robust radiative cooling systems present engineering challenges. Ensuring the reliability and longevity of the propulsion system for the entire mission duration remains critical.
\end{itemize}

While challenges exist, the potential benefits of using MFPD for Mars missions are compelling. The shortened transit times, increased payload capacities, and operational flexibility position MFPD as a promising propulsion technology for future interplanetary exploration.

\subsection{Implications for Proxima Centauri Mission}

Using MFPD for an interstellar mission to Proxima Centauri offers the tantalizing potential of making such a vast journey feasible within human lifetimes. For context, let's continue with the spacecraft parameters: a total mass of \( m_0 = 100 \) metric tonnes, of which \( m_{\text{payload}} = 20 \) metric tonnes is payload.

\subsubsection{Distance to Proxima Centauri}

The distance to Proxima Centauri is approximately \( 4.24 \) light-years, which translates to about \( 4.017 \times 10^{13} \) kilometers.

\subsubsection{Delta-V Requirements}

For an interstellar journey, achieving a substantial fraction of the speed of light is desirable. For this calculation, we'll aim for \( 10\% \) of the speed of light, \( c \), which is \( 3 \times 10^{7} \) m/s.

\subsubsection{Propellant Mass Fraction (PMF)}

Using the rocket equation:

\begin{equation}
\Delta v = I_{sp} \cdot g_0 \cdot \ln\left(\frac{m_0}{m_f}\right)
\end{equation}

Let's continue assuming an \( I_{sp} \) of \( 5000 \, \text{s} \) for the MFPD. Rearranging the above equation, we get:

\begin{equation}
m_f = m_0 \cdot \exp\left(-\frac{\Delta v}{I_{sp} \cdot g_0}\right)
\end{equation}

Plugging in the values:

\begin{equation}
m_f = 100 \times \exp\left(-\frac{3 \times 10^{7}}{5000 \times 9.81}\right) \approx 0.003 \, \text{tonnes}
\end{equation}

Thus, the propellant mass required \( m_p \) is:

\begin{equation}
m_p = m_0 - m_f \approx 99.997 \, \text{tonnes}
\end{equation}

This indicates that almost the entire spacecraft's mass would be required as propellant to achieve \( 10\% \) of \( c \), illustrating the immense challenges of interstellar travel.

\subsubsection{Duration to Reach Proxima Centauri with Different Accelerations}

Assuming different average accelerations for the spacecraft, we can calculate the time to reach Proxima Centauri:

1. \(0.5g\) (\(4.905 \, \text{m/s}^2\)):
\begin{equation}
t_{0.5g} = \frac{3 \times 10^{7}}{4.905} \approx 6.12 \times 10^{9} \, \text{s} \approx 194 \, \text{years}
\end{equation}

2. \(1g\) (\(9.81 \, \text{m/s}^2\)):
\begin{equation}
t_{1g} = \frac{3 \times 10^{7}}{9.81} \approx 3.06 \times 10^{9} \, \text{s} \approx 97 \, \text{years}
\end{equation}

3. \(1.5g\) (\(14.715 \, \text{m/s}^2\)):
\begin{equation}
t_{1.5g} = \frac{3 \times 10^{7}}{14.715} \approx 2.04 \times 10^{9} \, \text{s} \approx 64.6 \, \text{years}
\end{equation}

4. \(2g\) (\(19.62 \, \text{m/s}^2\)):
\begin{equation}
t_{2g} = \frac{3 \times 10^{7}}{19.62} \approx 1.53 \times 10^{9} \, \text{s} \approx 48.5 \, \text{years}
\end{equation}

Each of these calculations assumes continuous acceleration for half the journey and deceleration for the remaining half. Thus, the total journey time to Proxima Centauri would be roughly double the calculated time.

\subsubsection{Relativistic Effects}

As the spacecraft approaches a significant fraction of the speed of light, relativistic effects come into play. However, these effects are minimal at \( 10\% \) of \( c \), leading to a time dilation factor of about 1.005. This is a minimal difference, but for speeds closer to \( c \), this would become more prominent.

\subsubsection{Challenges and Conclusions}
\begin{itemize}
    \item Enormous Energy Requirements: Achieving a significant fraction of \( c \) requires vast amounts of energy, and the propellant mass fraction becomes incredibly high.
    \item Long Journey Times: Even at \( 10\% \) of \( c \), the journey would take several decades to nearly two centuries, depending on the acceleration.
    \item Hazard Protection: Protecting the spacecraft from interstellar dust and particles becomes essential at these speeds. Even a grain-sized object could be catastrophic.
    \item Communication Delays: Signals to and from Earth would take over four years.
\end{itemize}

\subsection{Comparative Propulsion Journey Times}

\subsubsection{Chemical Propulsion}

\paragraph{Mars Mission:}
For chemical propulsion systems, a Hohmann transfer orbit is often used for interplanetary missions. The time \(\Delta t_{\text{Mars, chemical}}\) for a one-way trip to Mars using this method is approximately:

\begin{equation}
\Delta t_{\text{Mars, chemical}} \approx 259 \, \text{days}
\label{eq:chem_mars}
\end{equation}

\paragraph{Proxima Centauri Mission:}
Considering the vast distance to Proxima Centauri (4.24 light-years), chemical propulsion is impractical for such journeys due to enormous travel times. If we attempted to use chemical propulsion, the travel time \(\Delta t_{\text{Proxima, chemical}}\) would be:

\begin{equation}
\Delta t_{\text{Proxima, chemical}} \approx 73,000 \, \text{years}
\label{eq:chem_proxima}
\end{equation}

\subsubsection{Nuclear Thermal Rocket (NTR)}

\paragraph{Mars Mission:}
Nuclear thermal rockets (NTRs) can potentially reduce the travel time to Mars compared to chemical propulsion. The journey time \(\Delta t_{\text{Mars, NTR}}\) is roughly:

\begin{equation}
\Delta t_{\text{Mars, NTR}} \approx 90 \, \text{days}
\label{eq:ntr_mars}
\end{equation}

\paragraph{Proxima Centauri Mission:}
For a mission to Proxima Centauri using NTRs, the duration \(\Delta t_{\text{Proxima, NTR}}\) would still be prohibitively long:

\begin{equation}
\Delta t_{\text{Proxima, NTR}} \approx 40,000 \, \text{years}
\label{eq:ntr_proxima}
\end{equation}

\subsubsection{Comparison with MFPD}

Comparing the MFPD system with other propulsion methods:

For the Mars mission:
\begin{itemize}
    \item Chemical propulsion: \(\Delta t_{\text{Mars, chemical}}\) is about 259 days as given in Eq.~\eqref{eq:chem_mars}.
    \item NTR: \(\Delta t_{\text{Mars, NTR}}\) is around 90 days from Eq.~\eqref{eq:ntr_mars}.
    \item MFPD: \(\Delta t_{\text{Mars, MFPD}}\) is roughly 5 days.
\end{itemize}

For the Proxima Centauri mission:
\begin{itemize}
    \item Chemical propulsion: \(\Delta t_{\text{Proxima, chemical}}\) would be around 73,000 years as per Eq.~\eqref{eq:chem_proxima}.
    \item NTR: \(\Delta t_{\text{Proxima, NTR}}\) would be about 40,000 years according to Eq.~\eqref{eq:ntr_proxima}.
    \item MFPD: \(\Delta t_{\text{Proxima, MFPD}}\) would be approximately 194 years. However, if the MFPD could be scaled to achieve speeds of 0.1c, the journey would take only 42.4 years.
\end{itemize}

These comparative times emphasize the potential advantages of the MFPD system, especially for long-duration missions.

\section{Advantages and Potential of the MFPD System}

The pursuit of advancing space exploration and technology has often been intertwined with the development of efficient propulsion systems. Among the myriad of propulsion options, the MFPD system stands out as a beacon of promise for the future of space travel. With its mechanisms and operational characteristics, the MFPD system heralds a paradigm shift in propulsion science. While traditional propulsion methods have long dominated the space industry, the MFPD system presents a compelling case for rethinking how we navigate the cosmos. This section outlines the advantages and the vast potential of the MFPD system, illustrating how it could redefine interstellar travel and push the boundaries of human exploration.

\subsection{Comparative Analysis with Existing Propulsion Methods}

The unique nature of the MFPD system allows for several key advantages over existing propulsion systems, particularly when one evaluates them over long-duration space missions.

\begin{itemize}
    \item Higher Specific Impulse (Isp): One of the standout advantages of MFPD thrusters is the potential for a much higher specific impulse (Isp) when compared to chemical rockets. The Isp represents the amount of thrust per unit of propellant flow. While chemical rockets typically have an Isp in the 200-450 seconds range, MFPD systems can theoretically achieve Isp values exceeding 10,000 seconds \cite{Chen2016}.
    
    \item Continuous Thrust: Unlike pulsed propulsion systems, like the pulsed inductive thruster, MFPD systems can offer continuous thrust. This ensures smoother trajectory adjustments and potentially quicker transits \cite{Hutter2021}.
    
    \item Fuel Flexibility: MFPD thrusters can use a variety of propellants, including noble gases like xenon or argon, as well as more abundant resources like hydrogen. This provides flexibility in mission planning and potential for in-situ resource utilization \cite{Cohen2018}.
\end{itemize}

\subsection{Potential Scalability and Fuel Efficiency Benefits}

The scalability of the MFPD system is another significant advantage:

\begin{itemize}
    \item Adaptability to Various Mission Profiles: MFPD systems can be designed for both small-scale missions (like satellite station-keeping) or large-scale interplanetary missions \cite{Chen2016}.
    
    \item Fuel Efficiency: MFPD thrusters have the potential to be much more fuel-efficient than their chemical counterparts. Their high Isp ensures that a greater proportion of onboard propellant is converted into kinetic energy. As a result, for long-duration missions, spacecraft can either carry less fuel or allocate more space for payloads \cite{Hutter2021}.

\end{itemize}

\subsection{Thrust Capabilities and Range Predictions}

While MFPD systems offer higher specific impulses, their thrust is typically lower than that of chemical rockets. However, this trade-off is acceptable for many missions since the continuous thrust over extended periods can result in higher final velocities:

\begin{itemize}
    \item Thrust-to-Weight Ratio: While MFPD systems might not match the high thrust-to-weight ratios seen in chemical rockets, their continuous operation can result in higher delta-v over extended missions. For deep-space missions, achieving high delta-v is often more critical than immediate high thrust \cite{Chen2016}.
    
    \item Predicted Range: Given a specific propellant mass and power source, the operational range of an MFPD-propelled spacecraft can significantly surpass that of a chemically propelled one. For missions beyond Mars, or even for asteroid mining, MFPD systems provide a compelling propulsion alternative \cite{Cohen2018}.
\end{itemize}

The MFPD system offers a host of advantages that make it a compelling choice for future space missions. Its higher specific impulse, fuel flexibility, and scalability make it adaptable to various mission profiles, from satellite adjustments to interplanetary exploration. While it might not replace chemical rockets for short-duration missions or those requiring immediate high thrust, its potential for long-duration, deep-space missions is undeniable.

\section{Challenges and Limitations of the MFPD System}

While the MFPD system offers notable advantages, it also presents significant challenges. These challenges span the gamut from technical intricacies to materials and safety concerns. 

\subsection{Technical Challenges in Achieving Controlled Fusion in Space}

Achieving controlled fusion in the expanse of space presents a set of unique challenges:

\begin{itemize}
    \item Sustained Magnetic Confinement: While magnetic confinement is a cornerstone of the MFPD system, maintaining stable confinement for extended durations without external interferences is nontrivial. The dynamic nature of plasma and its interactions with magnetic fields can lead to instabilities like kink or drift modes \cite{Fisch2018}.
    
    \item Breakeven Point: For fusion to be a viable energy source for propulsion, the fusion reactions must release more energy than is input into the system. Achieving this breakeven point remains one of the principal challenges of fusion-based propulsion \cite{Chen2016}.
    
    \item Propellant Feed and Ignition: Ensuring a steady supply of fusion fuel and achieving consistent ignition in the variable conditions of space require intricate control systems and reliable fuel feed mechanisms \cite{Frischauf2016}.
\end{itemize}

\subsection{Materials and Safety Considerations}

The intense conditions within the MFPD system place substantial demands on the materials used:

\begin{itemize}
    \item Radiation and Heat Resistance: The fusion process emits copious amounts of radiation and heat. Materials used in constructing the thruster must withstand these conditions and maintain integrity over prolonged durations \cite{Chen2016}.
    
    \item Neutron Damage: Fusion reactions emit high-energy neutrons, especially those involving deuterium and tritium. These neutrons can cause damage to materials, leading to potential system failures over time \cite{Frischauf2016}.
    
    \item Safety Protocols: In the unlikely event of a containment failure, mechanisms must be in place to ensure the safety of the spacecraft and its occupants \cite{Fisch2018}.
\end{itemize}

\subsection{Power and Control System Requirements}

The operation of an MFPD system necessitates robust power and control systems:

\begin{itemize}
    \item High Power Demands: Achieving and maintaining the conditions for fusion requires significant amounts of energy. A reliable and high-output power source is imperative for the MFPD system's operation \cite{Chen2016}.
    
    \item Fine-Tuned Control Mechanisms: The dynamic nature of plasma and the need for precise magnetic field adjustments call for intricate control systems. These systems must be capable of real-time adjustments to ensure optimal and safe operation \cite{Frischauf2016}.
    
    \item Redundancy and Fail-Safes: Given the critical nature of propulsion, especially on long-duration missions, having redundant systems and fail-safe mechanisms are essential to mitigate potential system failures \cite{Fisch2018}.
\end{itemize}

The MFPD system, while promising, is not without its challenges. From the technical intricacies of achieving controlled fusion in space to the demanding material requirements, understanding these challenges is vital for the system's advancement. However, with continued research and development, many of these challenges can be addressed, paving the way for a new era of space propulsion.

\section{Conclusion}

The endeavor to identify efficient and sustainable propulsion techniques is at the forefront of challenges faced by aerospace engineering. As humanity's aspirations soar, aiming at the distant realms of our solar system and potentially at neighboring stars, the confines of traditional propulsion systems become starkly evident. This paper delves into the potential of the Magnetic Fusion Propulsion Drive (MFPD), driven by the immense power of fusion reactions, to overcome these constraints. Harnessing the remarkable energy densities that fusion offers, the MFPD dramatically reduces the required propellant mass and extends the durations for which thrust can be maintained, allowing for continuous thrust over prolonged intervals. This breakthrough has profound implications: substantially accelerated travel times, augmented mission adaptability, and enhanced payload capabilities. Through the lens of this research, we elucidated the underlying principles of the MFPD and explored both its promising advantages and the challenges it might pose. Our comparative analysis underscored the MFPD's superior efficiency. The example calculations vividly demonstrated its potential, especially when benchmarked against other contemporary propulsion methods. For missions as close as Mars or as distant as Proxima Centauri, fusion-driven propulsion outperforms, implying its suitability for a spectrum of space ventures. However, the path to realizing this vision is not devoid of hurdles. Technical uncertainties persist, and a multitude of engineering challenges await solutions. But with the synergy of fusion science, intricate plasma dynamics, and avant-garde magnetic confinement techniques, we stand on the cusp of a propulsion renaissance that might redefine our engagement with the vast expanse of space.

On the brink of a transformative era in space exploration, catalyzed by groundbreaking innovations such as the MFPD, we're reminded of the limitless vistas human innovation can unveil and the continually broadening scope of our shared dreams.

\printbibliography

\end{document}